# Estimates of electron correlation based on density expansions

Jerry L. Whitten
Department of Chemistry
North Carolina State University
Raleigh, NC 27695 USA
email: whitten@ncsu.edu

*Key words: many-electron electronic structure methods, electron correlation methods, configuration interaction*

**Abstract**

Methods for estimating the correlation energy of molecules and other electronic systems are discussed based on the assumption that the correlation energy can be partitioned between atomic regions. In one method, the electron density is expanded in terms of atomic contributions using rigorous electron repulsion bounds, and, in a second method, correlation contributions are associated with basis function pairs. The methods do not consider the detailed nature of localized excitations, but instead define a correlation energy per electron factor that that is unique to a specific atom. The correlation factors are basis function dependent and are determined by from configuration interaction calculations on diatomic and hydride molecules. The correlation energy estimates are compared with the results of high-level configuration interaction calculations for a test set of twenty-seven molecules representing a wide range of bonding environments (average error of 2.6%). An extension based on truncated CI calculations in which d- and hydrogen p-type functions are eliminated from the virtual space combined with estimates of dynamical correlation contributions using atomic correlation factors is discussed and applied to the dissociation of several molecules.

**I. Introduction**

The description of many-electron systems by configuration interaction (CI) is a useful way to describe complex systems providing the problem can be reduced to a manageable size. There is a vast literature on ways to do this ranging from perturbation methods that generate configurations and evaluate energies efficiently to methods for partitioning large systems into localized electronic subspaces or ways to balance errors in systems that are being compared.[1-9] Relatively few configurations are required to dissociate molecules correctly or to create proper spin states, but dynamical correlation effects, particularly those associated with angular correlation, require higher spherical harmonic basis functions and this leads to a rapid increase in number of interacting configurations. Finding more efficient ways to treat large systems by coupled-cluster[10-11] and multireference methods[12-14] and methods that use non-orthogonal molecular orbitals[15] are continuing research topics in the quest to find increasing accurate descriptions of ground and excited states of molecules and materials.

In the present work, simple methods for estimating the correlation energy of molecules based on density expansions are explored. If estimates are sufficiently accurate, as is found to be the case, the correlation estimate could be for the entire molecule, or for a portion of a system that is less important, reserving rigorous configuration interaction for a region of primary importance such as a reaction site on a surface of a solid or in a large molecule.

## II. Scope of the present work

In this work, we carry out high-level configuration interaction (CI) calculations on molecules and look for electron correlation contributions that appear to be nearly invariant and transferable from one system to another. The objective is to obtain estimates of the correlation energy from properties of the single-determinant wavefunction by expanding the electron density. We also investigate dynamical correlation contributions separately and develop estimates based on density expansions which when combined with static correlation contributions obtained from smaller configuration interaction expansions give accurate total correlation energies.

Both approaches make use of an ansatz involving the probability of finding an electron at a point *(x,y,z)* in space

$$E^{corr} = \int \gamma(x,y,z) \rho(x,y,z) dv$$

where $\rho(x,y,z)$ is the electron density and $\gamma$ is a correlation factor that varies within the molecule. The Hohenberg-Kohn theorem would allow $\gamma$ to be expressed as a functional of the density,[16] however, we wish to develop an argument in an even simpler direction. The density can be decomposed into regional or atomic contributions and the interesting question is whether $\gamma(x, y, z)$ can be represented simply for such regions. To motivate the argument, we begin with a single-determinant solution of an *N*-electron molecule or system, for example, a self-consistent-field (SCF) solution or a wavefunction predicted by suitable potentials such as described in Ref. 17. In the present work, wavefunctions are constructed from flexible basis sets defined in the Appendix. The multi-reference configuration interaction method used to generate many-determinant wavefunctions is also described in the Appendix.[7] We now perform a unitary transformation of the occupied molecular orbitals (and separately the unoccupied orbitals) to localize orbitals about a given atom in the molecule or particle. Such a transformation can be carried out in many ways:[18-22] by overlap maximization with orbitals on a site, by exchange maximization, or simply by defining a large nuclear charge on the atomic site of interest, setting other nuclear charges zero. Each localization methods produces qualitatively the same set of localized orbitals that describe how electrons can pack into the spatial region around the selected site using occupied orbitals from the initial *N*-electron wavefunction. For the present analysis, we use the last localization method and find the SCF solution for an eight-electron system defined by this single site potential (the 1s orbitals are constrained to be the same as in the initial single-determinant wavefunction). The Hamiltonian is

$$H = \sum_{i}^{N=8} (T_i - \frac{Z}{r_i}) + \sum_{i<j}^{N=8} r_{ij}^{-1}$$

If the localization were about a hydrogen site, the subspace would have 2 electrons. We return now to the correct Hamiltonian of the *N*-electron system, one that has the proper nuclear charges, and carry out a CI calculation on the 8-electron subspace (or 2e for hydrogen) embedded in the remainder of the electron distribution defined by the localization transformation. At the single-determinant level, because the transformation is orthogonal, the total wavefunction remains the same. The electrons outside the subspace are present but constitute a fixed "core" of electrons that also includes the 1s-electrons. All virtual orbitals are included in the calculation. We interpret the energy lowering as the correlation energy of the 8-electron system associated with the atomic site (or 2e system for hydrogen). The localized orbitals correspond to $\sigma$ and $\pi$ bonds and nonbonding orbitals depending on the specific atomic site and nearest neighbors of the site.

In a large molecule, one could carry out such calculations on each site. Since bonds involve shared basis functions on different nuclei, the CI energies from the individual 8- or 2-electron calculations are not additive. Removing redundancies and renormalizing to the correct total number of electrons could provide a route to the total correlation energy of the molecule. Other investigators have used CI methods with considerable success to calculate local energy decompositions and fragment correlation energies.[23-24]

However, we shall not pursue such explicit methods here, but instead focus on the correlation energy of the individual 8- and 2-electron electron systems. Specifically, we want to know how much the 8- and 2-electron subsystem correlation energies vary for different bonding situations. Carbon, because of the many types of bonds that can be formed, is a case where the variations should be most pronounced. Table 1 give results obtained for a series of hydrocarbons and other molecules. We note that the correlation energy of subsystems defined by same atomic site does not vary greatly even though the hybridization and ligands change, e.g., for C the correlation energy per electron is $0.0211 \pm 0.002$. Similar small variations are found for different N and O sites. These results suggest an approximate correlation conservation principle for a given site whereby a larger correlation energy associated with one bonding direction must be accompanied by smaller correlation contributions in other directions. Such a balance can be attributed to the required orthogonality of different virtual orbitals that have large contributions to a specific site. We consider, however, the present results primarily as motivation for a more quantitative approach described in the next section.

**Table 1.** Configuration interaction calculations on 8- or 2-electron subspaces defined by orbital localization about different atomic sites in molecues. Occupied and virtual scf molecular orbitals are transformed separately; thus, the single-determinant total wavefunction and energy of the molecule are invariant. The correlation energy of the subspace embedded in the remainder of the electron distribution is defined as Eci-Escf.

|  | energy[a] SCF 1-det | loc site | energy subspace CI | corr E | corr E/per e |
|---|---|---|---|---|---|
| **ch4** | -40.2038 | c | -40.3755 | -0.1717 | -0.0215 |
|  |  | h | -40.2304 | -0.0265 | -0.0133 |
| **ethane** | -79.2027 | c | -79.3678 | -0.1651 | -0.0206 |
|  |  | h | -79.2290 | -0.0262 | -0.0131 |
| **ethylene** | -78.0525 | c | -78.2376 | -0.1851 | -0.0231 |
|  |  | h | -78.0788 | -0.0264 | -0.0132 |
| **acetylene** | -76.8385 | c | -76.9971 | -0.1586 | -0.0198 |
|  |  | h | -76.8666 | -0.0281 | -0.0141 |
| **benzene** | -230.7513 | c | -230.9209 | -0.1696 | -0.0212 |
|  |  | h | -230.7768 | -0.0254 | -0.0127 |
| **h2o** | -76.0476 | o | -76.2648 | -0.2172 | -0.0272 |
|  |  | h | -76.0740 | -0.0263 | -0.0132 |
| **nh3** | -56.2073 | n | -56.4035 | -0.1962 | -0.0245 |
|  |  | h | -56.2329 | -0.0256 | -0.0128 |
| **glycine** (nh2-ch2-cooh) | -282.9045 | c1 | -283.0632 | -0.1587 | -0.0198 |
|  |  | c2 | -283.0575 | -0.1530 | -0.0191 |
|  |  | n1 | -283.0980 | -0.1935 | -0.0242 |
|  |  | o1 | -283.1325 | -0.2280 | -0.0285 |
|  |  | o2 | -283.1211 | -0.2165 | -0.0271 |
|  |  | h1 | -282.9277 | -0.0232 | -0.0116 |
|  |  | h2 | -282.9284 | -0.0239 | -0.0119 |
|  |  | h3 | -282.9285 | -0.0239 | -0.0120 |
|  |  | h4 | -282.9302 | -0.0257 | -0.0128 |
|  |  | h5 | -282.9303 | -0.0257 | -0.0129 |

[a] Energies are in hartrees.

## III. Correlation via density expansions

The results of the previous section suggest that it shold be possible to partition molecular correlation into contributions associated with atomic sites. We return to the ansatz,

$$E^{corr} = \int \gamma(x,y,z)\rho(x,y,z)dv$$

and consider two questions: the representation of $\rho$ and choice of $\gamma$. First, we consider density expansions based on a rigorous 2-particle error bound, [25]

$$<\rho(1) - \rho'(1) | r_{12}^{-1} | \rho(2) - \rho'(2)> \geq 0$$

where $\rho$ is a single-determinant (SCF) density defined by occupied molecular orbitals, $\varphi_p$, and basis functions, $f_i$,

$$\rho = \sum_p \varphi_p \varphi_p = \sum_{i,j} w_{ij} f_i f_j$$

and $\rho'$ is a proposed approximation

$$\rho' = \sum_{i,j}^{i,j \in M} \lambda_{ij} f_i f_j \qquad (i, j \text{ on same atomic site M})$$

The coefficients $\lambda_{ij}$ are chosen to minimize the error bound and then renormalized to give the exact number of electrons.

Expansions based on minimizing the above error bound have been used for many purposes (see, for example, Refs. 25-28) and even a simplification of the approximate expansion to include only single site basis functions can be remarkably accurate as illustrated in Table 2 for a representative set of molecules. Density expansion errors for the full set of 27 molecules investigated in this work are depicted in Fig. 1 $\varepsilon = <\rho(1) - \rho'(1) | r_{12}^{-1} | \rho(2) - \rho'(2)> \geq 0$  The plot gives the expansion error, in percent, $100\varepsilon / <\rho(1) | r_{12}^{-1} | \rho(2)>$ for each molecule investigated.

**Table 2.** Expansion of electron densities based on minimization of the rigorous bound $\varepsilon = <\rho(1) - \rho'(1) | r_{12}^{-1} | \rho(2) - \rho'(2)> \geq 0$ where $\rho$ is the exact SCF density and $\rho'$ is an expansion containing only basis functions on the same site (see text). Values in parentheses are for an approximate expansion using coefficients from a Mulliken approximation for basis function products on different nuclei.

| | $<\rho(1) | r_{12}^{-1} | \rho(2)>$ | $<\rho'(1) | r_{12}^{-1} | \rho'(2)>$ | $\varepsilon$ (error) | % error[a] |
|---|---|---|---|---|
| **ethylene** | 70.3429 | 70.3346 (70.3133) | 0.0082 | 0.0117 |
| **acetylene** | 60.6273 | 60.6262 (60.5913) | 0.0011 | 0.0018 |
| **benzene** | 312.0811 | 312.0632 (311.9993) | 0.0179 | 0.0057 |
| **nc4h5(ring)** | 257.8693 | 257.8517 (257.7397) | 0.0176 | 0.0068 |
| **c6h5-nh2** | 406.4195 | 406.3978 (406.3241) | 0.0217 | 0.0053 |
| **glycine** | 315.7115 | 315.6888 (315.6155) | 0.0227 | 0.0072 |
| **c6h5-cooh** | 406.4195 | 406.3978 (406.3241) | 0.0217 | 0.0053 |
| **h2o** | 46.6980 | 46.6951 -46.6730 | 0.0029 | 0.0062 |
| **fhco** | 166.6740 | 166.6585 (166.6092) | 0.0155 | 0.0093 |

[a] The % error $= 100\varepsilon / <\rho(1) | r_{12}^{-1} | \rho(2)>$; energies are in hartrees.

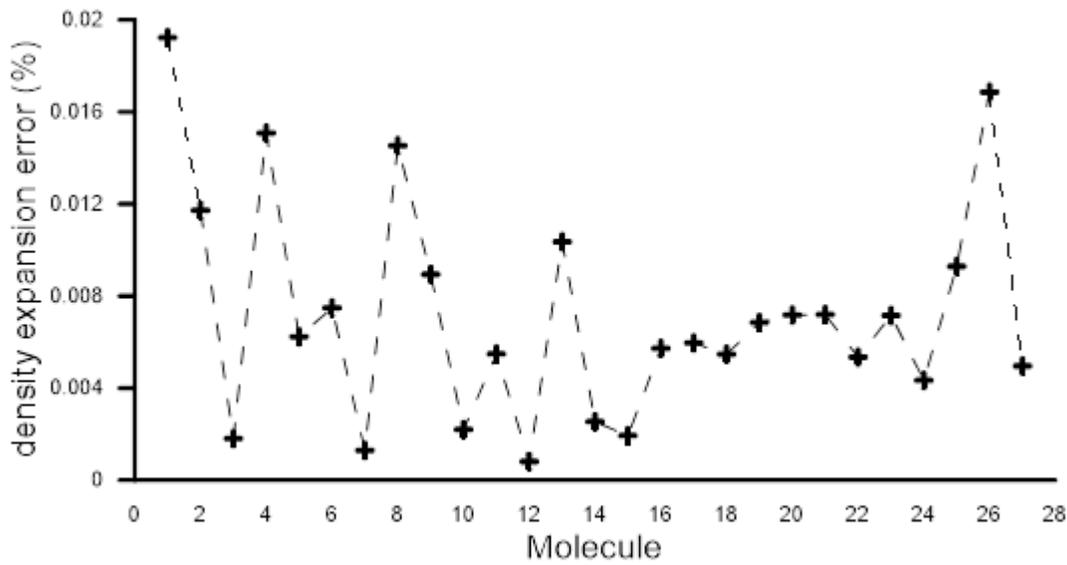

| | |
|---|---|
| 1 ch4 | 15 no |
| 2 c2h4 | 16 c6h6 |
| 3 c2h2 | 17 c4h4n2 |
| 4 c2h6 | 18 c5h5n |
| 5 h2o | 19 nc4h5 |
| 6 h2co | 20 nc4h4 |
| 7 co | 21 glycine |
| 8 c2 | 22 c6h5-nh2 |
| 9 o2 | 23 c5h5-cooh |
| 10 hf | 24 c6h5-cooh |
| 11 f2 | 25 fhco |
| 12 n2 | 26 c2f2h2 |
| 13 nh3 | 27 c6h5-f |
| 14 hcn | |

**Fig. 1. Density expansion based on electron repulsion bound.**

We now replace $\rho$ by the expansion $\rho'$

$$E^{corr} = \int \gamma(x, y, z) \rho'(x, y, z) dv$$

and make a *major* simplifying assumption that $\gamma$ depends only on the atomic site to give

$$E^{corr} = \int \gamma_M \sum_M \sum_{i,j}^{i,j \in M} \lambda_{ij} f_i f_j \, dv = \sum_M \sum_{i,j}^{i,j \in M} \gamma_M \lambda_{ij} <f_i | f_j> = \sum_M^{sites} \gamma_M p_M$$

where $p_M$ is the electron density or population associated with site M as determined by the $\rho'$ expansion. The choice of a single value of $\gamma_M$ for a given site suppresses detail about the origin of local excitations and assumes that the correlation energy associated with a site is conserved.

Values for $\gamma_M$ are determined from calculations on diatomic molecules $M_2$,

$$\gamma_M = (E^{CI} - E^{SCF}) / N_{M_2}$$

where $N_{M_2}$ is the number of electrons for the diatomic molecule minus the four 1s electrons not included in the CI. Table 3 lists the values for $\gamma_M$ obtained for the atoms considered in the present work, C, N, O, F and H; the diatomic molecule reference state is the lowest energy $^1\Sigma$ state except for $O_2$ where the state is $^3\Sigma$. The numerical value for $\gamma_M$ depends of course on the basis set. In the present work, first row atoms are described by a set of near Hartree-Fock atomic orbitals expanded as Gaussian functions plus an additional set of two-term 2s' and 2p' and two-term optimized 3d functions; hydrogens are described by a 1s, 1s' and optimized 2p functions. Thus, the present basis should give a good description of polarization and angular correlation effects in molecular systems. Basis functions details are included in Appendix. If the basis is changed significantly such that the correlation energy changes, the $\gamma_M$ values in the Table 3 may not be suitable and new values should be determined by repeating the diatomic molecule calculations. Also included in Table 3 are slightly refined values for $\gamma_M$ that correspond to the average values obtained for the diatomic molecule and the hydrides: $C_2$, $CH_4$; $N_2$, $NH_3$; $O_2$, $H_2O$; $F_2$, HF. We note that the optimum values for the diatomic and hydride systems differ only slightly for a given atom.

**Table 3**. Correlation factors (energies per electron) for diatomic and hydride molecules determined from density expansions based on the electron repulsion bound.

| atom | total correlation energy |  |  |
|---|---|---|---|
|  | diatomic | hydride | avg |
| F | 0.031462 | 0.029202 | 0.030332 |
| O | 0.032242 | 0.030116 | 0.031179 |
| N | 0.032430 | 0.029770 | 0.031100 |
| C | 0.028428 | 0.027600 | 0.028014 |
| H | 0.014000 | 0.014000 | 0.014000 |

We now consider a set of molecules representing different bonding environments and determine a correlation energy estimate based on the $\rho$ expansion and the diatomic-hydride average $\gamma$ values reported in Table 3. Energies are compared with a high-level multi-reference calculation on each molecule using the full basis (except for the 1s core electrons). The CI method is described in the Appendix. Correlation energy estimates for selected molecules are reported in Table 4, and, in Fig. 2, errors are given for the entire set of 27 molecules investigated. In both Table 4 and Fig. 2, the estimates are found to be in fairly good agreement ($\pm 5\%$) with the total correlation energy calculated by CI.

**Table 4.** Correlation energies from multi-reference CI calculations and estimates obtained from density expansions for selected molecules. Results for the full data set of 27 molecules are shown in Figure 2.

|  | SCF (1-det)[a] | CI (exact)[b] | Correlation energy | Correlation estimate | Error (%)[c] |
|---|---|---|---|---|---|
| **ethylene** | -78.0525 | -78.3518 | -0.2993 | -0.2968 | -0.85 |
| **acetylene** | -76.8385 | -77.1207 | -0.2822 | -0.2768 | -1.92 |
| **benzene** | -230.7513 | -231.5263 | -0.7750 | -0.7699 | -0.66 |
| **nc4h5 (ring)** | -208.8523 | -209.5435 | -0.6912 | -0.6783 | -1.87 |
| **c6h5-nh2** | -285.7960 | -286.7290 | -0.9330 | -0.9530 | 2.15 |
| **glycine (nh2-ch2-cooh)** | -282.9045 | -283.7218 | -0.8173 | -0.8394 | 2.71 |
| **c6h5-cooh** | -418.3925 | -419.6198 | -1.2274 | -1.2603 | 2.68 |
| **h2o** | -76.0476 | -76.2671 | -0.2194 | -0.2265 | 3.22 |
| **fhco** | -212.7663 | -213.2878 | -0.5215 | -0.5291 | 1.45 |

[a] Energies are in hartrees. [b] The energy of the multi-reference CI (see text) is referred to as "exact."

[c] % error=100(estimate-exact)/exact

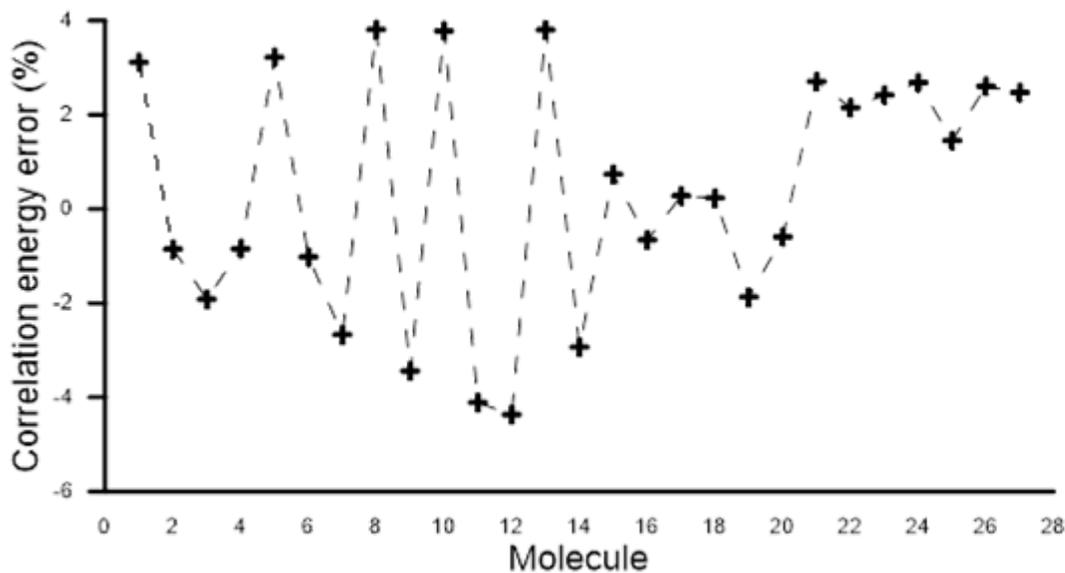

**Fig. 2. Correlation energy errors for estimates based on density expansions.** The black symbols are for errors in total correlation energy using only correlation factors and the atomic population determined by the density expansion. The energy of the multi-reference CI (see text) using the full basis is referred to as "exact." The % error=100(estimate-exact)/exact.

### IV. Correlation by partitioning of basis function contributions

Instead of introducing approximate density expansions, we consider an alternative approach based on the association of a correlation contribution with each basis function pair, $f_i f_j$, defined as $\frac{1}{2}(\gamma_i + \gamma_j)$ where $\gamma_i$ is determined by the atomic site of the basis function, and similarly for $\gamma_j$. If basis functions belong to the same site, M, the contribution is $\gamma_M$; if the sites are different, M and N, the contribution is $\frac{1}{2}(\gamma_M + \gamma_N)$. Then

$$E^{corr} = \int \gamma(x, y, z) \rho(x, y, z) dv$$

where, as before, $\rho$ is the single-determinant density

$$\rho = \sum_p \varphi_p \varphi_p = \sum_p \sum_{i,j} c_{pi} c_{pj} f_i f_j = \sum_{i,j} d_{ij} f_i f_j$$

And the total correlation contribution becomes

$$E^{corr} = \sum_{i,j} d_{ij} \tfrac{1}{2}(\gamma_i + \gamma_j) < f_i | f_j > = \sum_M^{sites} \gamma_M P_M .$$

Except for the definition of $P_M$, the expression is the same as in the density expansion based on the electron repulsion bound. If the density were fixed, we note that $P_M$ is the same as would have been obtained if the density $\rho$ were expanded using the Mulliken approximation. As seen in Table 2, the quality of the electron repulsion bound is reduced if Mulliken coefficients were used for the approximating density $\rho'$. However, in the present approach the quality of the correlation estimate depends not on the density expansion but on the validity of partitioning the basis function pair contribution. Correlation factors, $\gamma_M$, determined from calculations on diatomic and hydride molecules, as in the previous section, are reported in Table 5. The values are remarkably close to those reported in Table 3 from density expansions.

**Table 5**. Correlation factors (energies per electron) for diatomic and hydride molecules determined by partitioning basis function pair contributions (see text). Only the hydride values depend on the expansion method. The $\gamma_M$ are given for the total correlation energy and also for dynamical contributions only. The latter are used with CI expansions that exclude d and Hp functions from the virtual space (see text)

| | $\gamma_M$ for total correlation energy | | | $\gamma_M'$ for dynamical correlation only | | |
|---|---|---|---|---|---|---|
| atom | diatomic | hydride | avg | diatomic | hydride | avg |
| F | 0.031462 | 0.029260 | 0.030361 | 0.010857 | 0.011495 | 0.011176 |
| O | 0.032242 | 0.030190 | 0.031216 | 0.011094 | 0.012370 | 0.011732 |
| N | 0.032430 | 0.029850 | 0.031140 | 0.010360 | 0.012340 | 0.011350 |
| C | 0.028428 | 0.027320 | 0.027874 | 0.010361 | 0.010495 | 0.010428 |
| H | 0.014000 | 0.014000 | 0.014000 | 0.005000 | 0.005000 | 0.005000 |

Results are reported for selected molecules in Table 6 for the full set of 27 molecules in Fig. 3.

**Table 6.** Correlation energies from multi-reference CI calculations and estimates obtained by partitioning basis function pair contributions. Results for the full data set of 27 molecules are given in Fig. 3.

|  | SCF (1-det)[a] | CI (exact)[b] | Correlation energy | Correlation estimate | Error (%)[c] |
|---|---|---|---|---|---|
| **ethylene** | -78.0525 | -78.3518 | -0.2993 | -0.2946 | -1.56 |
| **acetylene** | -76.8385 | -77.1207 | -0.2822 | -0.2608 | -7.60[d] |
| **benzene** | -230.7513 | -231.5263 | -0.7750 | -0.7790 | 0.52 |
| **nc4h5 (ring)** | -208.8523 | -209.5435 | -0.6912 | -0.6918 | 0.09 |
| **c6h5-nh2** | -285.7960 | -286.7290 | -0.9330 | -0.9595 | 2.84 |
| **glycine (nh2-ch2-cooh)** | -282.9045 | -283.7218 | -0.8173 | -0.8537 | 4.45 |
| **c6h5-cooh** | -418.3925 | -419.6198 | -1.2274 | -1.2716 | 3.60 |
| **h2o** | -76.0476 | -76.2671 | -0.2194 | -0.2263 | 3.15 |
| **fhco** | -212.7663 | -213.2878 | -0.5215 | -0.5331 | 2.23 |

[a] Energies are in hartrees.  [b] The energy of the multi-reference CI (see text) is referred to as "exact."

[c] % error=100(estimate-exact)/exact

[d] The relatively large error for acetylene is due to the importance of $\pi_x \pi_y$ excitations; the error is reduced in Table 7.

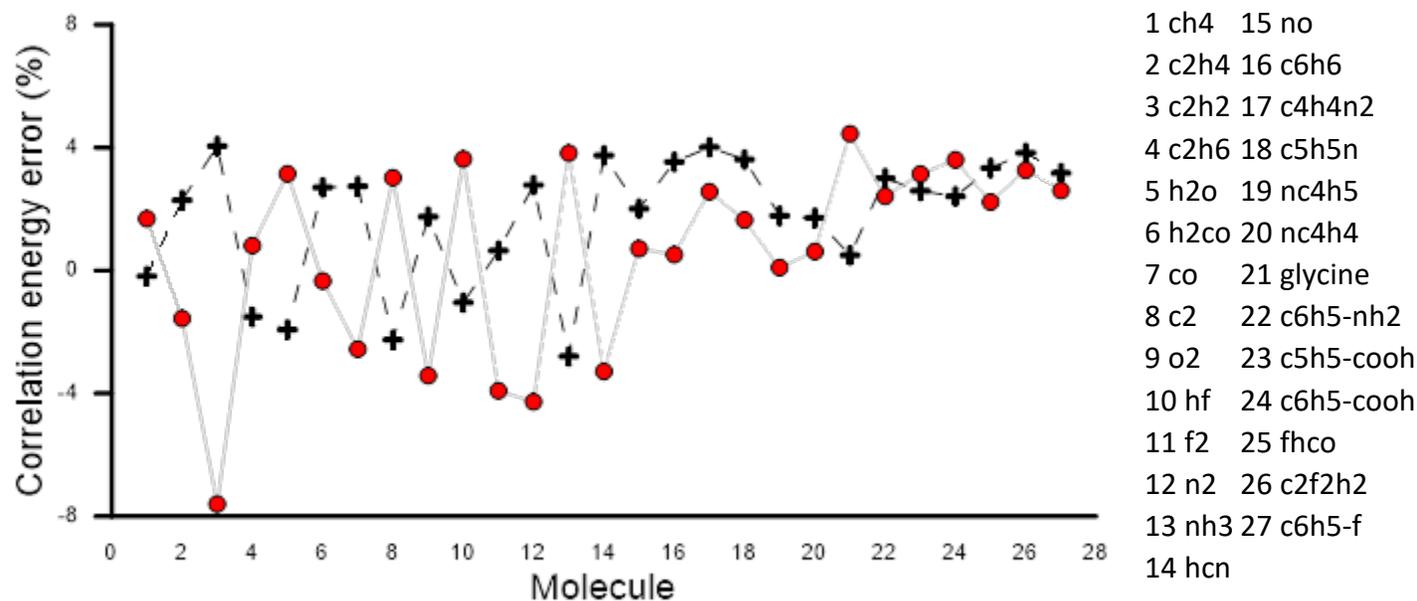

**Fig. 3. Correlation energy errors based on a partitioning of the basis function pair contribution.** The black symbols are for errors in total correlation energy calculated using only correlation factors. The red symbols are errors based on correlation factors for the dynamical correlation contribution plus the correlation energy contribution calculated explicitly by a CI calculation that excludes d-Hp virtual orbitals. The energy of the multi-reference CI (see text) using the full basis is referred to as "exact." The % error=100(estimate-exact)/exact.

As noted earlier, the rapid growth in number of configurations as molecules increase in size is in large part due to the higher spherical harmonic functions needed for dynamical correlation effects such as angular correlation. It would be useful if these contributions could be included more simply and the results of the previous section suggest that this should be possible. First, site specific contributions, $\gamma_M'$, are determined for each atom where the prime denotes dynamical contributions only, and not the total correlation energy. The procedure used is as follows:

a) An SCF calculation on the molecule of interest is carried out using the full basis. A virtual space is created by removing the d-type and hydrogen p-type functions (referred to subsequently as d-Hp functions). This can be done either by localization or simply by carrying out a SCF calculation for the virtual space with d-Hp basis functions removed. The virtual molecular orbitals are orthogonalized to the occupied SCF orbitals and to other virtual orbitals.

b) A CI calculation on the system with the set of occupied molecular orbitals (including d-Hp contributions) and the reduced set of virtual orbitals (excluding d-Hp functions except as required by orthogonality) is carried out to obtain a CI energy, $E'$. We refer to the difference $E^{dyn} = E^{full-CI} - E'$ as the dynamical correlation energy. This is the energy to be captured by $\gamma_M'$ where for a diatomic molecule. $\gamma_M' = E_{M_2}^{dyn} / N_{M_2}$ where $N_{M_2}$ is the number of electrons excluding the four 1s electrons.

Proceeding in this way avoids confusing the polarization and correlation roles of the d-Hp functions since the polarization effects are included in the 1-det SCF calculation. Values for $\gamma_M'$ are reported in Table 5.

To estimate the correlation energy of a molecule or other system, an SCF calculation is carried out using the full basis. The occupied molecular orbitals and the reduced set of orbitals for the virtual space (excluding d-Hp basis functions as noted above) are then used as a basis for a CI calculation. For a given total electron density, the estimate of the total correlation energy becomes

$$E^{corr} = E^{CI} + \sum_M^{sites} \gamma_M' P_M$$

Implicit in the argument is the existence in principle of a d-Hp virtual space around a site and that correlation contributions involving excitations to that space, although not carried out explicitly, are captured by $\gamma_M'$ and the population of the site,. Table 7 and Fig. 3 show that the assumptions work remarkably well. In contrast to the use of $\gamma_M$ to estimate the total correlation contribution, the present treatment can be applicable to excited electronic states and molecular dissociation since the CI allows for occupancy changes. In the last section, the dissociation of several molecules is examined. The CI calculations are simplified by this approach, but the treatment does not simplify the evaluation of integrals over basis functions or molecular orbitals except for reducing the number of the latter orbitals.

Comparing the correlation energy errors in Tables 6 and 7 and Fig. 3 with the earlier values based on the density expansion using the electron repulsion bound shows that both approaches have comparable errors. The average deviations are 2.2% for the density expansion and 2.6% in the latter case. Although conceptually different, both methods have the same structure for a fixed density expansion and both methods support the idea that correlation contributions can be partitioned into contributions associated with atomic sites in a system.

**Table 7.** Correlation energies from multi-reference CI calculations and dynamical correlation obtained by partitioning basis function pair contributions for selected molecules. The correlation estimate is the sum of the value calculated by a CI that excludes d-Hp virtual orbitals and the dynamical correlation estimate. Results for the full data set of 27 molecules are given in Figure 3.

| | SCF (1-det)[a] | CI (exact)[b] | Correlation energy | Correlation estimate | Error (%)[c] |
|---|---|---|---|---|---|
| **ethylene** | -78.0525 | -78.3518 | -0.2993 | -0.3062 | 2.29 |
| **acetylene** | -76.8385 | -77.1207 | -0.2822 | -0.2937 | 4.05 |
| **benzene** | -230.7513 | -231.5263 | -0.7750 | -0.8024 | 3.54 |
| **nc4h5 (ring)** | -208.8523 | -209.5435 | -0.6912 | -0.7035 | 1.78 |
| **c6h5-nh2** | -285.7960 | -286.7290 | -0.9330 | -0.9610 | 3.00 |
| **glycine (nh2-ch2-cooh)** | -282.9045 | -283.7218 | -0.8173 | -0.8214 | 0.49 |
| **c6h5-cooh** | -418.3925 | -419.6198 | -1.2274 | -1.2570 | 2.41 |
| **h2o** | -76.0476 | -76.2671 | -0.2194 | -0.2152 | -1.93 |
| **fhco** | -212.7663 | -213.2878 | -0.5215 | -0.5389 | 1.03 |

[a] Energies are in hartrees. [b] The energy of the multi-reference CI (see text) is referred to as "exact."
[c] % error=100(estimate-exact)/exact

The second method is extraordinarily simple and instead of working with the total density, correlation contributions can be introduced by simply modifying the one-electron potential energy integral over basis functions.

$$< f_i \,|\mathrm{V}|f_j > \quad \rightarrow \quad < f_i \,|\mathrm{V}|f_j > + \tfrac{1}{2} < f_i \,|f_j > (\gamma_M + \gamma_N)$$

Or, equivalently, we can define a modified Hamiltonian for the system as

$$H'' = H + \sum_i h_i''$$

where H is the exact Hamiltonian and $h_i''$ is a one electron operator that carries the correlation contribution, defined by its matrix elements

$$< f_i |h''|f_j > = \tfrac{1}{2} < f_i |f_j > (\gamma_M + \gamma_N).$$

The expectation value of $H''$ then includes the correlation energy estimate. The correlation contribution slightly affects the iterations of an SCF calculation transferring charge to the atom with larger $\gamma$. If there were no difference in values of $\gamma$ there would be no change in the Fock operator since for orthogonal orbitals occupied $\varphi_p$ and virtual $\varphi_q$ since $<\varphi_p|\gamma|\varphi_q> = \gamma <\varphi_p|\varphi_q> = 0$. The correlation energies reported in Tables 6-7 and Fig. 3 contain the self-consistent-field adjustment.

**Molecular Dissociation**

In this section, we consider the dissociation of molecules and treatments based on truncated CI calculations in which d-Hp virtual orbitals are excluded and dynamical correlation contributions are estimated by the partitioning of the basis function pair contribution. To probe the accuracy, the following dissociations are considered:

1) simple single bond, $H_2O \rightarrow OH + H$

2) complex single bond, $H_2N\text{-}H_2C\text{-}COOH \rightarrow H_2N\text{-}H_2C + COOH$

3) double bond, $H_2C=CH_2 \rightarrow H_2C + CH_2$ ($CH_2$ in high spin, S=1 state)

4) triplet state multiple bond, $O_2 \rightarrow O + O$ ( S=1 state for O and $O_2$ )

In Fig. 4., potential energy curves are calculated and compared with the "exact" CI result. The agreement is found to be quite good over the entire range of distances.

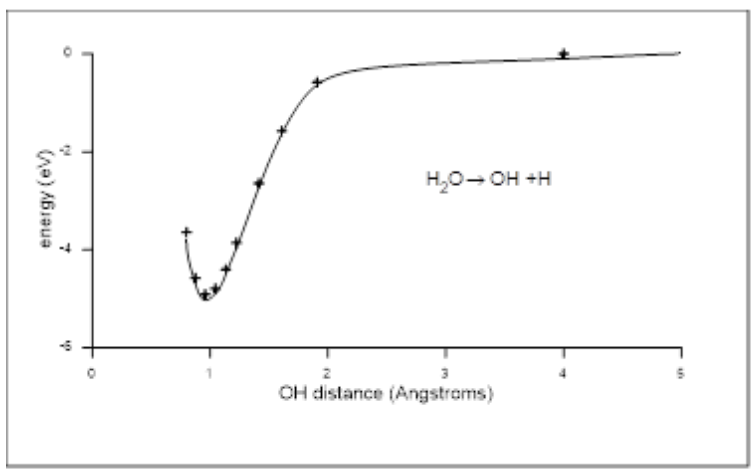
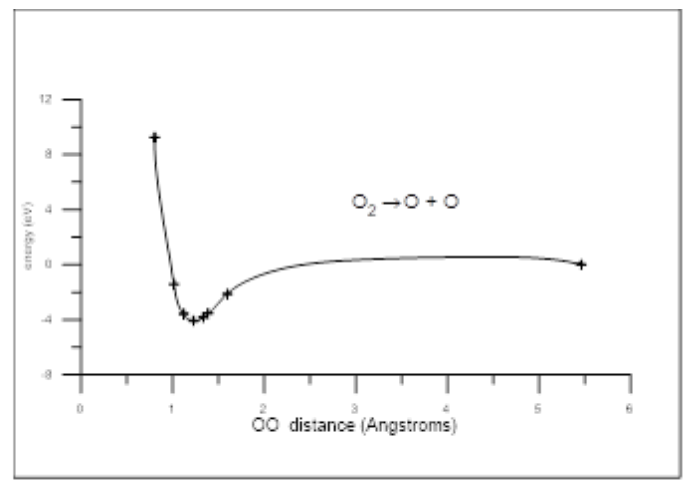
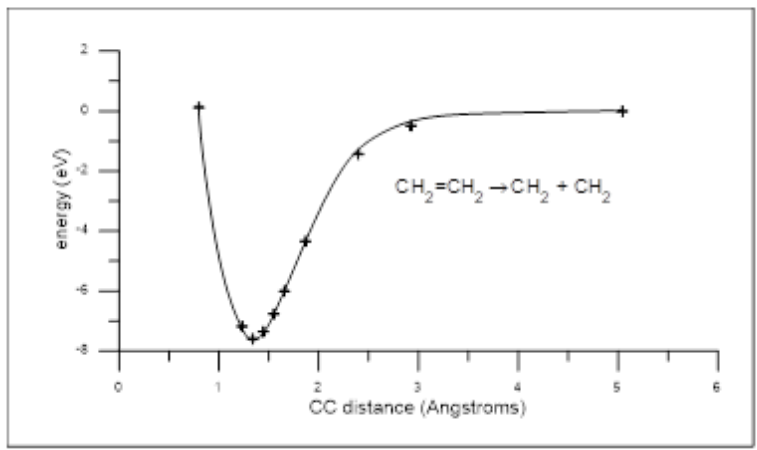
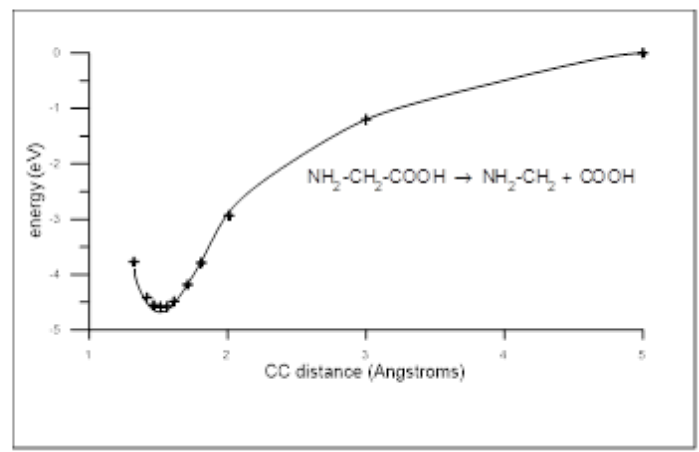

**Fig. 4. Dissociation of single and multiply bonded molecules.** The black crosses denote the energies obtained from truncated CI calculations with no d-Hp virtual orbitals plus the dynamical correlation energy estimate. The solid line is a cubic spline fit to the exact CI energies.

**Conclusions**

Two methods for estimating the correlation energy of molecules and other electronic systems are discussed based on the assumption that the correlation energy can be partitioned between atomic regions. In (1), the electron density is expanded in terms of atomic contributions using rigorous electron repulsion bounds, and, in (2), correlation contributions are associated with basis function pairs. The methods do not consider the detailed nature of localized excitations, but instead define a correlation energy per electron factor, $\gamma_M$, that that is unique to a specific atom $M$. Method (2) can be implemented simply by adding to the exact Hamiltonian a one-electron operator, $\sum_k h_k''$, that carries the correlation information; $h_i''$ is defined by its matrix elements $<f_i|h''|f_j> = \tfrac{1}{2}<f_i|f_j>(\gamma_M+\gamma_N)$ for a given basis where $M$ and $N$ are nuclei associated with $f_i$ and $f_j$. The correlation factors are basis function dependent and are determined by from configuration interaction calculations on diatomic and hydride molecules. Correlation energy estimates are compared with the results of high-level configuration interaction calculations on twenty-seven molecules representing a wide range of bonding environments (average error 2.6%). An extension based on smaller CI expansions in which the d- and hydrogen p-type functions are eliminated from the virtual space combined with estimates of dynamical correlation contributions using atomic correlation factors is discussed and applied to the dissociation of several molecules. The existence of nearly invariant atomic factors associated with localized correlation contributions suggests that the correlation contributions in regions around an atom including bonds and nonbonding electrons is approximately conserved even if the bonding environment changes when single or multiple bonds are formed. This can also be shown by starting with a single-determinant wavefunction for a given basis and localizing the molecular orbitals about individual nuclei in a system.

**Appendix**

**Basis set**

The basis for each atom is a near Hartree-Fock set of atomic orbitals plus extra two-component s- and p-type functions consisting of the two smaller exponent components of the atomic orbital; sets of two-component d and two-component p functions are added for first-row atoms and hydrogen, respectively. The latter d- and p-type functions were optimized by CI calculations on atoms. Orbitals are expanded as linear combinations of Gaussian functions: 1s(10), 2s(5), 2p(5), 2s′(2), 2p′(2), for C,N,O , 2p(6) for F, and 1s(4), s(1) for H where the number of Gaussian functions in each orbital is indicated in parentheses. No core potentials were used in the present calculations so that the predictive capability of the methods could be fully tested.

**Configuration interaction**

All calculations are carried out for the full electrostatic Hamiltonian of the system

$$H = \sum_i^N [-\tfrac{1}{2}\nabla_i^2 + \sum_k^Q -\frac{Z_k}{r_{ik}}] + \sum_{i<j}^N r_{ij}^{-1}$$

A single-determinant self-consistent-field (SCF) solution is obtained initially for each state of interest. Configuration interaction wavefunctions are constructed by multi-reference expansions,[7-8]

$$\Psi = \sum_k c_k (N!)^{-1/2} det(\chi_1^k \chi_2^k \ldots \chi_N^k) = \sum_k c_k \Phi_k$$

In all applications, the entire set of SCF orbitals is used to define the CI active space. Virtual orbitals are determined by a positive ion transformation to improve convergence. Single and double excitations from the single determinant SCF wavefunction, $\Phi_r$, creates a small CI expansion, $\Psi_r'$,

$$\Psi_r' = \Phi_r + \sum_{ijkl} \lambda_{ijkl} \Gamma_{ij \to kl} \Phi_r = \sum_m c_m \Phi_m$$

The configurations $\Phi_m$, are retained if the interaction with $\Phi_r$ satisfies a relatively large second order energy condition

$$\frac{|<\Phi_m|H|\Phi_r>|^2}{E_m - E_r + \lambda} \geq 10^{-4}$$

The description is then refined by generating a large CI expansion, $\Psi_r$ by single and double excitations from all important members of $\Psi_r'$ to obtain

$$\Psi_r = \Psi_r' + \sum_m \left[ \sum_{ik} \lambda_{ikm} \Gamma_{i \to k} \Phi_m + \sum_{ijkl} \lambda_{iklm} \Gamma_{ij \to kl} \Phi_m \right]$$

where $\Phi_m$ is a member of $\Psi_r'$ with coefficient $> 0.01$. Typically, $\Psi_r'$ contains 200-400 dets. We refer to this expansion as a multi-reference CI. The additional configurations are generated by identifying and retaining all configurations, $\Phi_m$, that interact with $\Psi_r'$ such that

$$\frac{|<\Phi_m|H|\Psi'_r>|^2}{E_m - E_r + \lambda} \geq 10^{-6}$$

For the molecules investigated, approximately $10^5$-$10^6$ determinants occur in the final CI expansion, and the expansion can contain single through quadruple excitations from an initial representation of the state $\Phi_r$. The contribution of determinants not explicitly included along with size consistency corrections are estimated by perturbation theory. The value of $\lambda$ is determined so that the second order perturbation energy matches the CI value if first order coefficients

$$c_m = \frac{-<\Phi_m|H|\Psi'_r>}{E_m - E_r + \lambda}$$

are used for determinants in the CI calculation.

**Acknowledgment**

Discussions with Professor Mike Whangbo are gratefully acknowledged.